# Characterization study of a broad-energy germanium detector at CJPL

ZENG Zhi[1]   MI Yuhao[1]   ZENG Ming[1]   MA Hao[1,*]   YUE Qian[1]   CHENG Jianping[1]
LI Junli[1]   QIU Rui[1]   ZHANG Hui[1]

[1] Key Laboratory of Particle and Radiation Imaging (Ministry of Education) and Department of Engineering Physics, Tsinghua University, Beijing 100084, China

*Corresponding author, *E-mail address:* mahao@tsinghua.edu.cn

**Abstract:** The ability of background discrimination using pulse shape discrimination (PSD) in broad-energy germanium (BEGe) detectors makes them as competitive candidates for neutrinoless double beta decay (0νββ) experiments. The measurements of key parameters for detector modeling in a commercial p-type BEGe detector are presented in this paper. Point-like sources were used to investigate the energy resolution and linearity of the detector. A cylindrical volume source was used for the efficiency calibration. With an assembled device for source positioning, a collimated $^{133}$Ba point-like source was used to scan the detector and investigate the active volume. A point-like source of $^{241}$Am was used to measure the dead layer thicknesses, which are approximately 0.17 mm on the front and 1.18 mm on the side. The described characterization method will play an important role in the 0νββ experiments with BEGe detectors at China JinPing underground Laboratory (CJPL) in the future.

## 1. Introduction

Broad-energy germanium (BEGe) detectors are competitive candidates for neutrinoless double beta decay (0νββ) experiments using germanium detectors. Benefiting from the unique electrode structure with a rather small inner electrode for signal output, BEGe detectors have a better performance of energy resolutions and a stronger pulse shape discrimination (PSD) power for single-site events (SSEs) and multi-site events (MSEs) than semi-coaxial germanium detectors[1,2]. This PSD power plays an important role in the background discrimination for 0νββ experiments. GERDA has used several BEGe detectors for 0νββ detection in Phase I and achieved the most sensitive result among all similar experiments using germanium detectors[3]. In the on-going Phase II, GERDA is using much more BEGe detectors to take full advantage of their good performances[4].

During the research of the PSD power of BEGe detectors, fine detector models in both Monte Carlo (MC) simulation and pulse shape simulation (PSS) must be established so as to evaluate the discrimination efficiency of a developed PSD method and study the features of pulse shapes induced by different physical events. In such detector models, the active volume size and the dead layer thickness are key parameters for obtaining accurate simulation results. Thus, it is necessary to extract specific values of these parameters through characterization of BEGe detectors. GERDA has devoted lots of energies to the characterization of BEGe detectors[5-7], while other researchers also performed some study[8,9]. In these previous work, investigation of the gamma spectrometry performance, extraction of key detector parameters and study of the pulse shape features were always concerned.

In the future, 0νββ experiments with BEGe detectors will be conducted at China JinPing underground Laboratory (CJPL). As the deepest underground laboratory in the world, CJPL has a rock overburden of 2400 m (about 6720 m w.e.)[10] and a muon flux of $2.0\times10^{-6}$ m$^{-2}$s$^{-1}$ [11]. As the preliminary study of BEGe detectors for the 0νββ experiments at CJPL, the characterization of a commercial BEGe detector at CJPL is presented in this work. The energy resolution and linearity of the detector were investigated and the efficiencies were calibrated. The active volume of the detector was scanned with a collimated $^{133}$Ba point-like source. The dead layer thicknesses on the front and side of the detector were measured with a $^{241}$Am point-like source.

## 2. Method and experiment

### 2.1. Detector specification

The detector under investigation is a commercial p-type BEGe detector (Model BE6530) produced by Canberra[12] and it has been stored at CJPL for about 4 years. A schematic view of the detector configuration is shown in Figure 1. The germanium crystal has a diameter of 91.1 mm and a height of 31.4 mm. The small boron-implanted p+ electrode is 13.5 mm in diameter and serves as the signal contact. The lithium-diffused n+ electrode covers most of the residual surface of the crystal, serves as the high-voltage contact and is separated from the p+ electrode by an annular groove. The crystal is held by a copper cup in a 1.6-mm-thick aluminum endcap and placed 8 mm from the front window. The front window is made of 0.6-mm-thick carbon composite to enhance the detection efficiencies of low-energy gamma rays that penetrate from the front. The recommended bias voltage is +4500 V.

The data acquisition system in this work involves a charge-sensitive pre-amplifier (Model 2002C), an integrated digital signal analyzer (Model DSA-LX) and the Genie-2000 software. The pre-amplifier is integrated with the detector and pre-amplifies the charge signal from the p+ electrode. The digital signal analyzer (DSA) integrates functions of the high-voltage module, main amplifier module and multi-channel analyzer module in an analog electronics chain. The DSA records the signal pulse shapes from the pre-amplifier with a fast sampling ADC (FADC), extracts their energy information through a firmware with the trapezoidal shaping algorithm and finally sends the information to the Genie-2000 software, which addresses the production and storage of energy spectra.

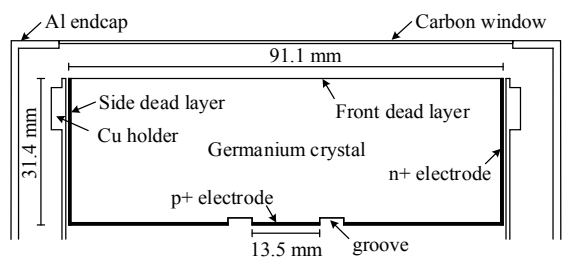

Figure 1 Schematic view of the BEGe detector.

### 2.2. Experimental procedure

Procedures of the measurements of the energy resolution, the linearity, the efficiencies, the active volume size and the dead layer thicknesses are described in detail as below:

(a) Energy resolution: point-like sources of $^{241}$Am, $^{133}$Ba, $^{137}$Cs, $^{60}$Co and $^{152}$Eu were located approximately 7 cm above the top surface of the detector to measure its energy resolution. As to the point-like sources, the radioactive materials are sealed as a dot (1 mm in diameter) in a thin plastic film (Φ 32 mm × 4 mm). At first, the $^{60}$Co source was measured alone to adjust the parameters (rise time & flat top) of the trapezoidal shaping algorithm of the DSA. Then the energy resolutions of more gamma peaks with energies between 59.5 keV-1408 keV from the above radioisotopes were obtained to study their dependency on the gamma peak energy.

(b) Linearity: the same sources as in Procedure (a) were measured and the peak locations of the corresponding gamma rays were recorded to check the linearity of the detector's response to the gamma ray energy deposition.

(c) Efficiency: a certified cylindrical volume source, which was obtained from UK's National Physical Laboratory (NPL) in a comparison measurement, made of filter medium and had more than 10 radioactive isotopes ($^{241}$Am, $^{57}$Co, $^{60}$Co, $^{88}$Y, $^{54}$Mn, etc), was placed on the top surface of the detector to obtain the dependency of the absolute detection efficiency on the gamma ray energy. As for a concerned gamma ray, the net peak count was calculated and used to deduce the corresponding detection efficiency, taking into consideration the measurement time, the emission intensity of the gamma ray and the source radioactivity.

(d) Active volume size: the same $^{133}$Ba source as in Procedure (a) was 1 mm collimated with a 1-cm-thick stainless-steel collimator and used to scan the BEGe detector along the diameter of the top surface as well as the height of the lateral surface with a 1 cm or 0.5 cm step. The collimator was located as close to the endcap surface as possible in case the enlightened areas of the crystal in adjacent steps interlap each other. The incident collimated gamma beams were perpendicular to the detector surfaces. The measurement was constant with time for each position in a certain scanning. The net peak count of the 81 keV gamma ray was continuously recorded during this process, and the size of the active volume was deduced from its variation.

(e) Dead layer thickness: the same $^{241}$Am source as in Procedure (a) was used to measure the front and side dead layer thicknesses. In a certain measurement, the source was placed at a fixed position and the full-energy-peak (FEP) detection efficiency of the 59.5 keV gamma ray was obtained experimentally (given the certified source activity of ~ 8290 Bq). Meanwhile, the FEP detection efficiency was also obtained through MC simulation based on GEANT4 with the identical setup as the experiment. The simulation was performed repeatedly and the thickness value of the corresponding dead layer in the detector model was changed every time, thus the dependency of the FEP detection efficiency on the dead layer thickness was obtained. This dependency relationship was fitted with an exponential function[13], and the real dead layer thickness was determined by interpolating the acquired function to the experimentally acquired FEP detection efficiency.

### 2.3. Assembled collimation device

To locate point-like sources at different positions around the detector, an assembled device made of stainless steel was set up to aid the characterization. Figure 2 shows a concept view of the manually operated device. It is composed of three main parts:

(a) Holding structure: an L-shaped holder that supports all other parts and keeps them stable.

(b) Position-fixing part: through the rotation of the central shaft and movement of the source holder, point-like sources can be located above the top surface or around the lateral surface of the detector with an error of 1 mm.

(c) Removable collimators: with the collimators, it is possible to scan the detector with collimated radiation beams in different directions.

Parts (a) and (b) can be disassembled into smaller parts for convenient transportation and storage.

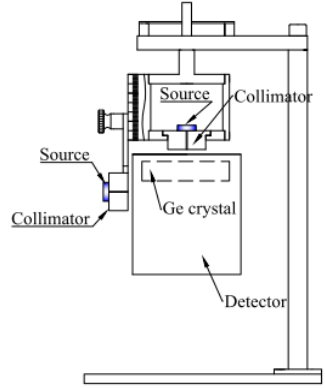

Figure 2 Concept view of the assembled collimation device for source positioning.

## 3. Results and discussion

### 3.1. Energy resolution

For germanium detectors, energy resolutions usually refer to the full width at half maximum (FWHM) of a gamma peak. With a smaller inner signal contact, BEGe detectors can achieve a lower electronics noise level and consequently a better energy resolution than semi-coaxial HPGe detectors[14]. In 0νββ experiments, good energy resolutions are important to narrow the region of interest, distinguish signals from background and improve the experiment sensitivity[15].

Figure 3 presents the variations in energy resolutions of the gamma rays from $^{60}$Co (1173 keV & 1332 keV) when the rise time and flat top of the trapezoidal shaping algorithm of the DSA were adjusted. In Figure 3 (a), the FWHM slightly fluctuated within 0.05 keV, while the rise time was fixed at 6 μs and the flat top was changing between 0.8 μs and 1.2 μs. In Figure 3 (b), with the flat top fixed at 1 μs, the FWHM quickly increased when the rise time was below 2 μs but fluctuated slightly within 0.06 keV between 2 μs and 10 μs. Finally, the flat top and rise time were fixed at 1 μs and 6 μs respectively for the subsequent measurements in this study. Based on further measurements with corresponding point-like sources described in Section 2.2, the FWHM was obtained and fitted as a function of the gamma peak energy (in keV) by[16]:

$$\text{FWHM} = a \times \sqrt{E} + b,$$

where $E$ is the gamma peak energy and $a$, $b$ are the coefficients. The result is shown in Figure 4. The FWHM of the 1.33 MeV peak is approximately 1.66 keV (~0.125%), which is outstanding among germanium detectors.

## 3.2. Linearity

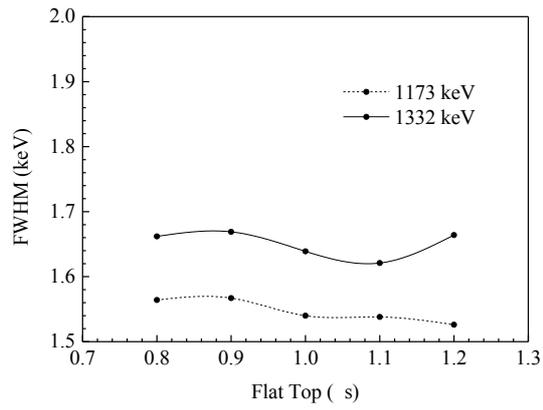

(a)

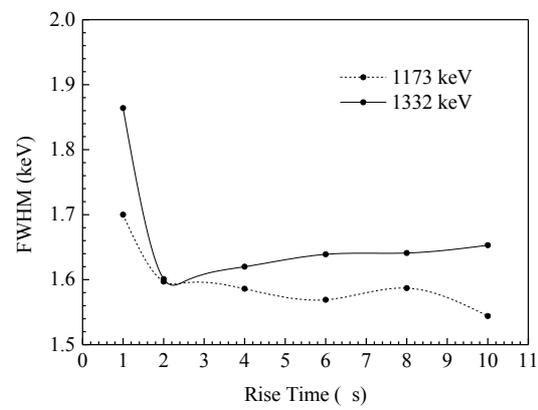

(b)

Figure 3 Energy resolutions of gamma rays from $^{60}$Co. In (a), the rise time was fixed at 6 μs; in (b), the flat top was fixed at 1 μs.

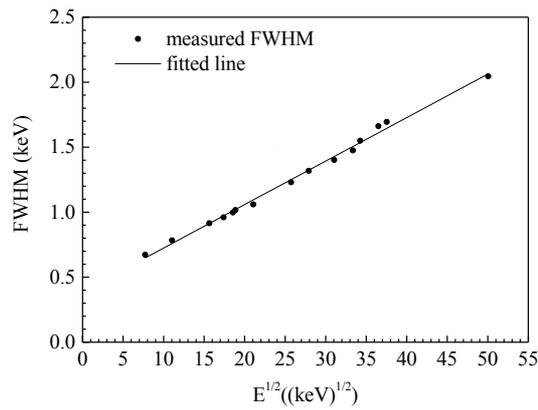

Figure 4 Energy resolutions of the BEGe detector. The fitted fuction is $\text{FWHM} = 0.033 \times \sqrt{E} + 0.39$ ($R^2 = 0.9949$).

## 3.2. Linearity

The linearity of the responses of germanium detectors to different energy depositions is usually excellent, which is reflected by the goodness when the energy calibration curve is fitted to a linear function. A good linearity leads to a clear discrimination of different physical events based on their energy information.

To study the energy response linearity of the BEGe detector, corresponding point-like sources described in Section 2.2 were measured. Considering that the gamma ray with the largest energy is the 1408 keV one from $^{152}$Eu, the summation peak that resulted from the coincidence effect of the two gamma rays of $^{60}$Co was also used to compensate for the possible deviation from linearity in the high-energy region during the fitting process. The linearity of the detector is notably good, as observed in Figure 5 where the energy is in keV.

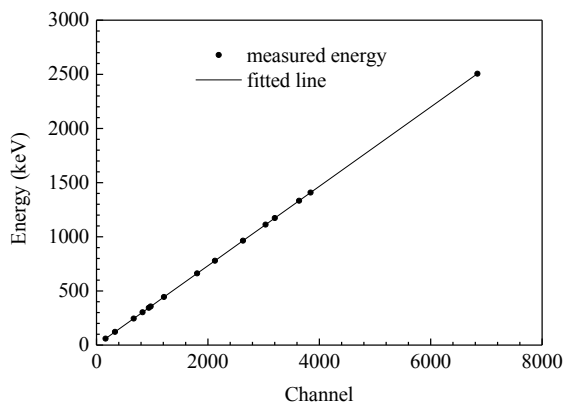

Figure 5 Linearity of the energy response of the BEGe detector. The fitted function is $E = 0.37 \times CH - 0.13$ ($R^2 = 1.0000$), where $E$ is the deposited energy in the detector and $CH$ is the channel number in the spectrum.

## 3.3. Efficiency

The absolute detection efficiencies of the BEGe detector were measured using the certified cylindrical volume source described in Section 2.2, and the involved gamma rays are listed in Table 1. Figure 6 shows the efficiency curve as a function of the gamma ray energy (in keV), which is fitted by[17]:
$\ln eff = a + b \times \ln E + c \times (\ln E)^2 + d \times (\ln E)^3$,
where *eff* is the absolute detection efficiency, $E$ is the gamma ray energy and a, b, c, d are the coefficients. The curve reaches its maximum at approximately 70 keV and decreases almost linearly above 200 keV in double logarithmic coordinates. Before the maximum point, gamma rays pass through the entrance window and the n+ electrode of the detector with an increasing probability as their energies increase; once they succeed in reaching the active volume of the detector, almost all of them are absorbed. After the maximum point, gamma rays are absorbed by the detector with a decreasing probability as their energies increase.

Table 1 Gamma rays used for the measurement of absolute detection efficiencies.

| Energy (keV) | Nuclide |
|---|---|
| 46.5 | $^{210}$Pb |
| 59.5 | $^{241}$Am |
| 88.0 | $^{109}$Cd |
| 122.1 | $^{57}$Co |
| 136.5 | $^{57}$Co |
| 165.9 | $^{139}$Ce |
| 391.7 | ¹¹³Sn |
| 661.7 | ¹³⁷Cs |
| 834.8 | ⁵⁴Mn |
| 898.0 | ⁸⁸Y |
| 1115.5 | ⁶⁵Zn |
| 1173.2 | ⁶⁰Co |
| 1332.5 | ⁶⁰Co |
| 1836.1 | ⁸⁸Y |

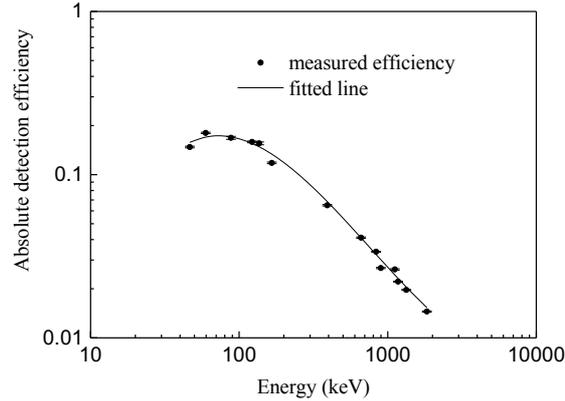

Figure 6 Absolute detection efficiencies of the BEGe detector. The efficiencies were calibrated using a cylindrical volume source. The fitted function is $\ln eff = -14.55 + 7.1 \times \ln E - 1.224 \times (\ln E)^2 + 0.0616 \times (\ln E)^3$ ($R^2 = 0.9955$).

### 3.4. Volume scanning

As previously described in Section 2.2, the measurements were conducted with the collimated ¹³³Ba source and we paid attention to the variations in the net peak count of the 81 keV gamma ray, which corresponds to the curves in Figure 7. Figure 7 (a) shows the result along the diameter of the top surface, and position 0 cm corresponds to the center of the surface. Ideally, the curve should have a flat central plateau and a sharp decline at the boundary of the germanium crystal. However, in our case, the net peak count was maximal at the center, slowly decreasing when the source moved outwards and quickly declined when the source arrived at the boundary. The difference is attributed to the incomplete collimation of the source with the 1-cm-thick collimator, of which the thickness was chosen as a result of the compromise considering the notably low activity (in the level of several kBq) of the source. When a thicker collimator was used, the collimation improved but the peak almost disappeared in the background. This incomplete collimation would influence the accuracy of the result a lot. If the diameter of the active volume is defined as the width with the net peak count above 50% of that at the center, the active volume here was determined to be approximately 87.7 ± 2 mm in diameter, whereas the diameter of the crystal from the manufacturer is 91.1 mm.

Figure 7 (b) shows the result vertically along the lateral surface, and position 0 cm corresponds to the same horizontal level as that of the center of the crystal. The curve does not behave as expected because of the incomplete collimation and the configuration around the crystal. To help understand the result and obtain the internal structure of the Al endcap, an X-ray image of the detector was generated, as shown in Figure 8. When the source moved along the upper part of the crystal, the net peak counts decreased because of the shielding of the thicker copper holder around this part of the crystal; when the source moved around the upper boundary of the crystal, the net peak count increased abnormally instead of decreasing because of the increasing detection probability of the uncollimated gamma rays through the large top surface of the crystal. On the contrary, the net peak counts increased when the source moved along the lower part of the crystal with the thinner copper holder around and continued decreasing at the boundary because of the shielding of the copper holder which covers the bottom surface of the crystal. Consequently, it is hard to deduce the thickness of the active volume from this result, and a well-collimated, strong source is necessary for improvements in the future.

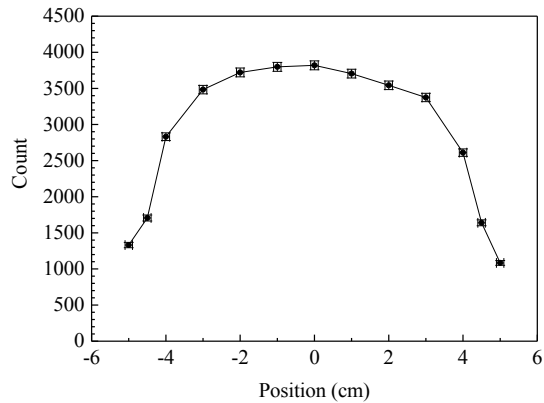

(a)

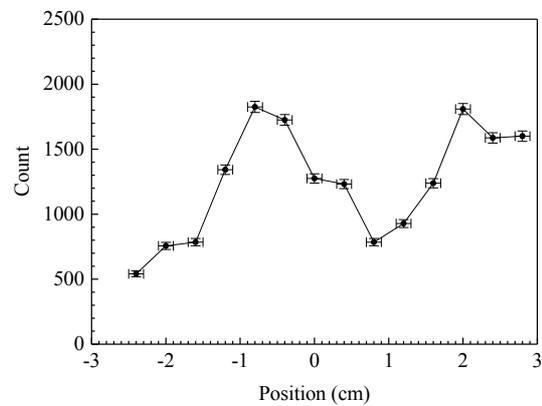

(b)

Figure 7 Net peak count of the 81 keV gamma peak as a function of the position of a collimated $^{133}$Ba source. (a) Scanning along the top surface of the detector; (b) scanning along the lateral surface.

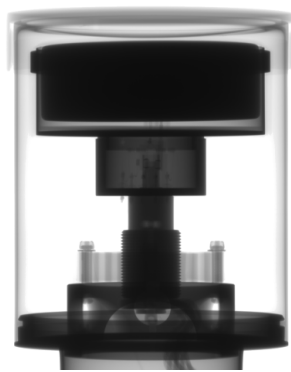

Figure 8 An X-ray image of the BEGe detector

### 3.5. Dead layer

Germanium detectors always have dead layers on the surface where electric fields are so weak that electrons and holes cannot be fully collected[17]. Therefore, if a radioactive particle deposits some energy in the dead layers, the total energy recorded by the detector will deviate from the actual energy deposition. For commercial BEGe detectors, typical dead layer thicknesses provided by the manufacturer are merely reference values instead of exclusive ones. Besides, dead layers may grow as time goes by[18], especially when the detector is stored without applying high voltage for a long time. The certified $^{241}$Am point-like source described in Section 2.2 was used for the measurements of dead layer thicknesses in this work.

The $^{241}$Am source was approximately 4 cm above the center of the top surface of the detector when we measured the front dead layer thickness, which corresponds to the energy spectrum in Figure 9. The 59.5 keV peak was fitted using the combination of a Gaussian function and a linear one. The net peak count and the FEP detection efficiency were derived based on the fitting result. The curve in Figure 10 derived from the simulation shows the dependency of the FEP detection efficiency on the front dead layer thickness, and the straight lines indicate the determined thickness value. The final result is 0.166 ± 0.011 (stat) ± 0.1 (syst) mm, whereas the typical value from the manufacturer is 0.3 μm. It is also worth mentioning that the standard systematic errors of all results of dead-layer thicknesses in Section 3.5 were estimated to be 0.1 mm, which is a conservative value covering the errors introduced during the processes of source positioning, fitting, simulation, etc.

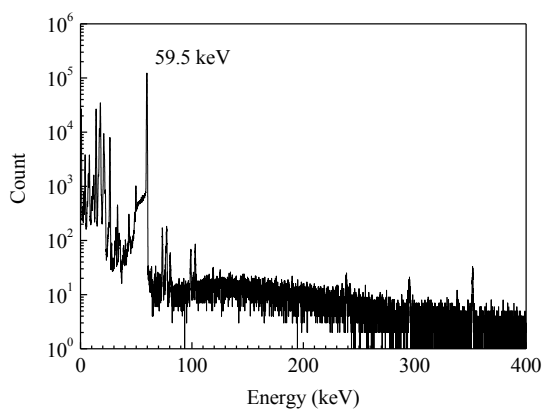

Figure 9 Energy spectrum of the $^{241}$Am source above the top surface of the detector.

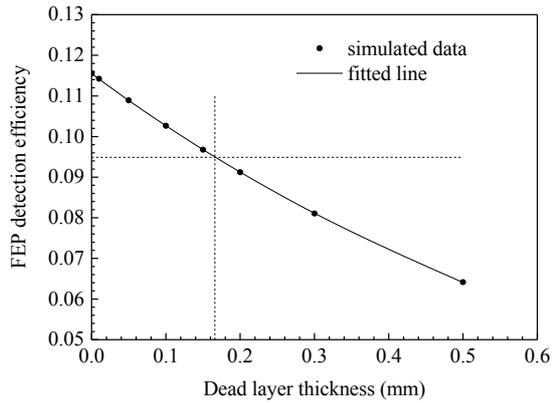

Figure 10 Dependency of the FEP detection efficiency of the 59.5 keV gamma ray on the front dead layer thickness, as obtained from the MC simulation. The dashed black lines indicate the experimental efficiency value and the corresponding dead layer thickness value evaluated from the fitted exponential function of the thickness dependency. The fitted function is $eff = 0.1155 \times e^{-1.178 \times t}$ ($R^2 = 1.0000$), where *eff* is the FEP detection efficiency and *t* is the front dead-layer thickness.

When the side dead layer thickness was measured, the $^{241}$Am source was located at two different rotation angles (referred to as the 0° direction and the 45° direction) respectively, on the same horizontal level as that of the center of the germanium crystal and approximately 3 cm away from the lateral surface of the detector. Similar analyses as that for the top measurement were conducted. The energy spectrum and the dependency of the FEP detection efficiency on the side dead layer thickness for the 0° direction are presented in Figure 11 and Figure 12 respectively, while those for the 45° direction are omitted. The determined dead layer thickness values for both directions are listed in Table 2, which fit well with each other and lead to the average of 1.179 mm compared to the typical value of 0.5 mm from the manufacturerTable 2.

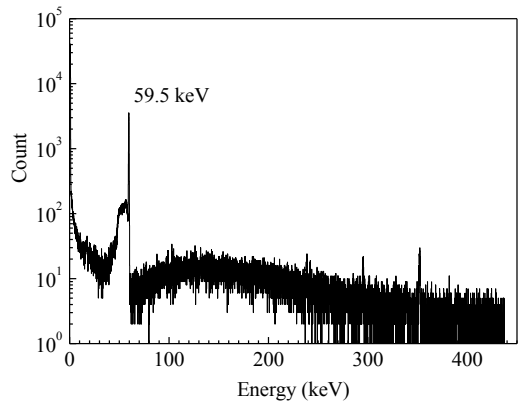

Figure 11 Energy spectrum of the $^{241}$Am source in the 0° direction.

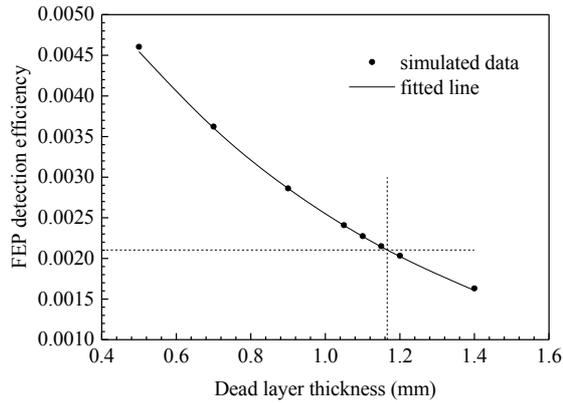

Figure 12 Dependency of the FEP detection efficiency of the 59.5 keV gamma ray on the side dead layer thickness, as obtained from the MC simulation. The dashed black lines indicate the experimental efficiency value and the corresponding dead layer thickness value evaluated from the fitted exponential function of the thickness dependency. The fitted function is $eff = 0.0081 \times e^{-1.156 \times t}$ ($R^2 = 0.9997$), where *eff* is the FEP detection efficiency and *t* is the side dead-layer thickness. The result corresponds to the $^{241}$Am source in the 0° direction.

The $^{133}$Ba point-like source described in Section 2.2 was also used to measure the side dead layer thickness at two different rotation angles, as a benchmark with the measurements using $^{241}$Am. The $^{133}$Ba source was located at the same positions as those of the $^{241}$Am source. The whole process was similar to that using $^{241}$Am, except that the ratio between the FEP detection efficiencies of two gamma rays of $^{133}$Ba was obtained in both the experiment and simulation. The FEP detection efficiency ratio between one gamma ray with a lower energy and another with a higher energy is sensitive to the dead layer thickness, thus it can also be used to determine the thickness value. Here the FEP detection efficiency ratios of the 81 keV gamma ray to the 276 keV, 303 keV, 356 keV and 384keV gamma rays respectively were calculated. The energy spectrum and the dependency of the FEP detection efficiency ratios on the side dead layer thickness for the 0° direction are presented in Figure 13 and Figure 14Figure 12 respectively, while those for the 45° direction are omitted. The determined dead layer thickness values from different pairs of gamma rays are also listed in Table 2, which all fit with the results from the $^{241}$Am source.

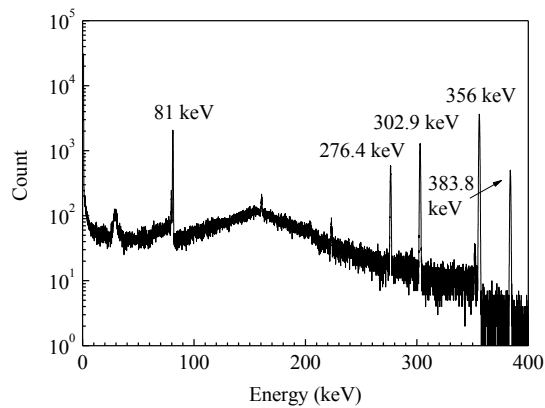

Figure 13 Energy spectrum of the $^{133}$Ba source in the 0° direction.

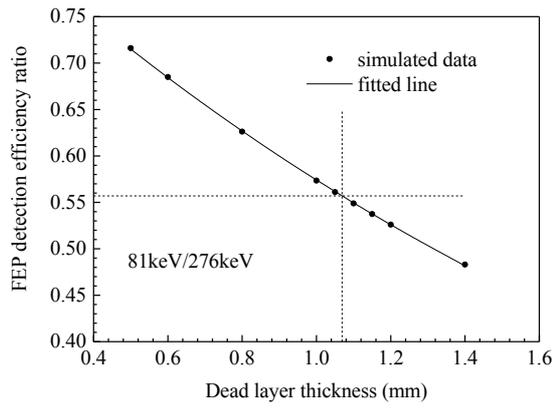

(a)

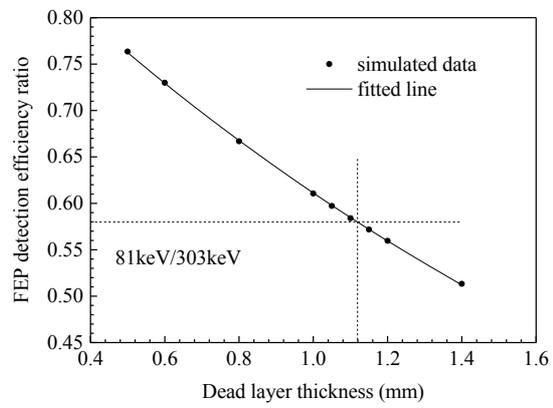

(b)

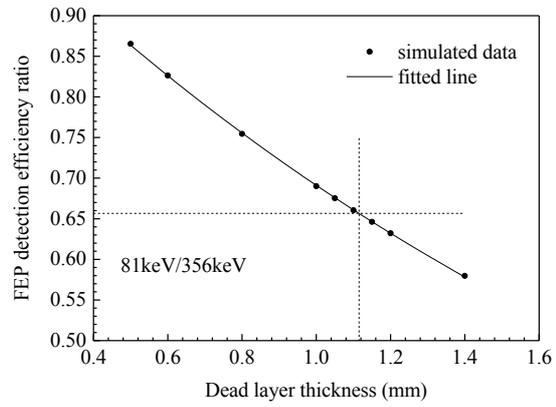

(c)

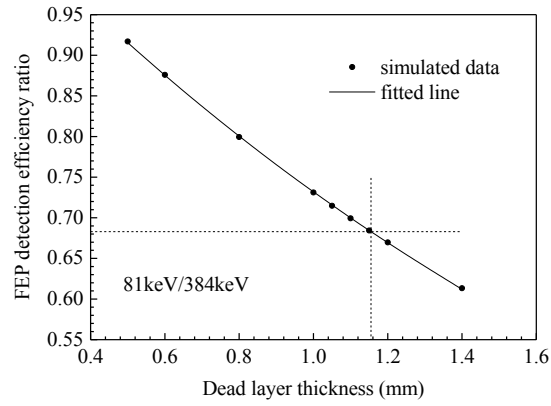

(d)

Figure 14 Dependencies of the FEP detection efficiency ratios among different gamma rays of the $^{133}$Ba source on the side dead-layer thickness, as obtained from the Monte Carlo simulation. The dashed black lines indicate the experimental ratio values and the corresponding dead-layer thickness values evaluated from the fitted exponential functions of thickness dependencies. The fitted functions are: (a) $r = 0.8906 \times e^{-0.4389 \times t}$ ($R^2 = 0.9999$), (b) $r = 0.9511 \times e^{-0.4422 \times t}$ ($R^2 = 0.9999$), (c) $r = 1.080 \times e^{-0.4459 \times t}$ ($R^2 = 0.9999$), (d) $r = 1.145 \times e^{-0.4474 \times t}$ ($R^2 = 0.9999$), where $r$ is the FEP detection efficiency ratio and $t$ is the side dead-layer thickness. All results correspond to the $^{133}$Ba source in the 0° direction.

Table 2 Measurement results of the side dead-layer thickness (units: mm).

|  |  | 0° | 45° |
|---|---|---|---|
| $^{241}$Am | 59.5 keV | 1.166 ± 0.011 (stat) ± 0.1 (syst) | 1.192 ± 0.012 (stat) ± 0.1 (syst) |
| $^{133}$Ba | 81 keV/276 keV | 1.069 ± 0.029 (stat) ± 0.1 (syst) | 1.122 ± 0.028 (stat) ± 0.1 (syst) |
|  | 81 keV/303 keV | 1.119 ± 0.021 (stat) ± 0.1 (syst) | 1.152 ± 0.022 (stat) ± 0.1 (syst) |
|  | 81 keV/356 keV | 1.115 ± 0.017 (stat) ± 0.1 (syst) | 1.191 ± 0.017 (stat) ± 0.1 (syst) |
|  | 81 keV/384 keV | 1.155 ± 0.028 (stat) ± 0.1 (syst) | 1.219 ± 0.028 (stat) ± 0.1 (syst) |

## 4. Summary

The characterization of a commercial BEGe detector using an assembled collimation device was conducted at CJPL, as the preliminary study for the 0νββ experiments in the future. The detector's gamma spectrometry performance is excellent: the energy resolution of the 1.33 MeV peak reaches 1.66 keV, the energy response linearity is perfect and the efficiency curve behaves as expected. The diameter of the active volume is approximately 87.7 mm resulting from the scanning process, whereas a good result of the thickness was not available due to the imperfect experiment conditions. The front and side dead layer thicknesses are approximately 0.17 mm and 1.18 mm respectively, which have increased compared to the typical values from the manufacturer. The determined dead layer thickness values will be applied to the detector models for MC simulation and PSS, whereas the active volume size will be further measured using a strong source with good collimation.

In future, the study on pulse shapes of the BEGe detector is planned to be performed. A systematic pulse shape analysis method will be established for BEGe detectors, which will contribute to the background discrimination in the prospective 0νββ experiments at CJPL.

## Acknowledgments

This work is partly supported by National Natural Science Foundation of China (No.11175099 &No. 11355001) and Tsinghua University Initiative Scientific Research Program (No.20151080354 & No.2014Z21016).